# Effects of friction and plastic deformation in shock-comminuted damaged rocks on impact heating

Kosuke Kurosawa[1]* and Hidenori Genda[2]

[1]Planetary Exploration Research Center, Chiba Institute of Technology, 2-17-1, Narashino, Tsudanuma, Chiba 275-0016, Japan

[2]Earth–Life Science Institute, Tokyo Institute of Technology, 2-12-1 Ookayama, Meguro-ku, Tokyo 152-8550, Japan

*Corresponding author
Kosuke Kurosawa, Ph.D.
Planetary Exploration Research Center, Chiba Institute of Technology
E-mail: kosuke.kurosawa@perc.it-chiba.ac.jp
Tel: +81-47-4782-0320
Fax: +81-47-4782-0372

**Abstract**
Hypervelocity impacts cause significant heating of planetary bodies. Such events are recorded by a reset of $^{40}$Ar–$^{36}$Ar ages and/or impact melts. Here, we investigate the influence of friction and plastic deformation in shock-generated comminuted rocks on the degree of impact heating using the iSALE shock-physics code. We demonstrate that conversion from kinetic to internal energy in the targets with strength occurs during pressure release, and additional heating becomes significant for low-velocity impacts (<10 km s$^{-1}$). This additional heat reduces the impact-velocity thresholds required to heat the targets with the 0.1 projectile mass to temperatures for the onset of Ar loss and melting from 8 and 10 km s$^{-1}$, respectively, for strengthless rocks to 2 and 6 km s$^{-1}$ for typical rocks. Our results suggest that the impact conditions required to produce the unique features caused by impact heating span a much wider range than previously thought.

# 1. Introduction

Collisions between two planetary bodies at speeds of several km s$^{-1}$ cause significant heating of their surface materials (e.g., Ahrens and O'Keefe, 1972), resulting in the generation of impact melts, which are among the most curious geological samples known (e.g., Grieve and Cintala, 1992). Since the degree of impact heating depends strongly on the impact velocity, detailed geochemical analyses of such heated samples allow us to characterize the impact environment in the solar system through its history. A recently proposed dynamical model of the early solar system predicts large-scale orbital migration of gas giants (e.g., Gomes et al., 2005), suggesting that the impact-velocity distributions on planetary bodies such as asteroids are significantly disturbed at the time of migration (e.g., Bottke et al., 2012). Such changes in the impact-velocity distribution might be recorded by the abundance of impact melts as a function of time (e.g., Marchi et al., 2013).

Understanding the impact velocities required for incipient and complete melting is essential to extract information about the impact-velocity distribution on planetary bodies. Ahrens and O'Keefe (1972) proposed that the entropy matching method could be used to quantify the velocity thresholds for incipient and complete melting by assuming that the shocked matter expands adiabatically to the reference volume; i.e., d$S$ = 0, where $S$ is the specific entropy. It is widely believed that this assumption is valid for impacts at relatively high velocities, because the strongly shocked matter would behave as a perfect fluid, and high thermal pressure leads to rapid adiabatic expansion. An advantage of the use of this method is that we only need to know the entropy in the peak-shock and reference states: we do not need to know the time dependence of thermodynamic quantities in phase space during pressure release.

This method has been widely used to obtain the peak pressure required for the incipient melting of various geological materials (e.g., Ahrens and O'Keefe, 1972; Pierazzo et al., 1997; Pierazzo et al., 2005; Stewart et al., 2008; Hamann et al., 2016). The corresponding impact velocities for the onset of melting can be estimated using the one-dimensional impedance matching method (e.g., Ahrens and O'Keefe, 1972). For example, the estimated velocity thresholds for granite and basalt are 5.1 km s$^{-1}$ (46 GPa) and 7.6 km s$^{-1}$ (96 GPa), respectively. Note that these estimates are based on the assumption of head-on collisions between objects composed of the same material.

Quintana et al. (2015) carried out two-dimensional impact simulations using the CTH code and reported that the material strength leads to a higher degree of shock-induced melting than the purely hydrodynamic (i.e., no material strength) case when the impact velocity is relatively low (<10 km s$^{-1}$). Kenkmann et al. (2013) reported the results of impact experiments under MEMIN (Multidisciplinary Experimental and Modeling Impact Research Network; Kenkmann et al., 2011); i.e., that the partial melting of iron meteorites, which are launched as projectiles, occurs at a much lower peak pressure (55 GPa) than the peak pressure estimated for incipient melting of iron predicted by the entropy-matching method (162 GPa). Such unexpected additional heating has not been recognized in laboratory experiments, possibly because the amount of permissible strain in uniaxial (one-dimensional) shock–recovery experiments is highly limited with respect to the three-dimensional deformation during natural impacts.

This changes the velocity thresholds for impact melting to somewhat lower values than previously thought. Although Quintana et al. (2015) did not clearly show how the material strength enhances impact melting and which physical parameters are important in the material strength model, their new insights could have a significant influence on decoding impact histories based on the occurrence of impact melts.

In this study, we address how material strength affects the degree of impact heating using the iSALE code (Amsden et al., 1980; Ivanov et al., 1997; Collins et al., 2004; Wünnemann et al., 2006), which has been widely distributed to academic users in the impact community. We consider a simple, well-established constitutive model to address the energy partitioning from the kinetic energy of the impactor to the internal energy of both projectile and target. In particular, we focus on the influence of strength in the shock-comminuted damaged rocks on the degree of shock heating.

## 2. Methods

We used the two-dimensional model included in the iSALE shock-physics code, known as the iSALE–Dellen model (Collins et al., 2016). The strength model for rocks (Ivanov et al., 1997; Collins et al., 2004) and the ANalytical Equation Of State (ANEOS; Thompson and Lauson, 1972) for dunite (Benz et al., 1989) were applied to both projectile and target. This EOS model is commonly used to approximate the bulk properties of chondritic materials (e.g., Johnson et al., 2015).

In the strength model, the yield strength is given by

$$Y = (1-D)Y_i + DY_d, \qquad (1)$$

where $Y_i$ and $Y_d$ are the yield strengths of intact and shock-comminuted damaged rock, respectively. A damage parameter, $D$, is introduced, which expresses the reduction in strength with increasing plastic strain, varying between $D = 0$ (intact rocks) and $D = 1$ (shock-comminuted damaged rock). Since the unconfined crushing strength of intact rocks (1–5 GPa) is much lower than the shock compression under typical impact conditions (higher than a few km/s), the damage parameter $D$ reaches unity immediately upon shock-wave arrival near the impact point. Therefore, the additional heating caused by $Y_i$ is not significant (see also Section 3). Thus, here we focus on the effects of $Y_d$ on additional heating. The yield strength $Y_d$ is well established and is known as the Coulomb friction law,

$$Y_d = Y_{coh} + \mu_{dam}P, \qquad (2)$$

where $Y_{coh}$, $\mu_{dam}$, and $P$ are the cohesion, internal friction and temporal pressure, respectively. The yield strength $Y_d$ is limited by the von Mises plastic limit, which is typically 1–5 GPa under strong compression. A detailed description of the pressure-dependent yield strength used in this study is presented in Supporting Information S1. Since the additional heating caused by $Y_{coh}$ is not significant (see Section 3), $\mu_{dam}$ was treated as a free parameter so as to systematically investigate the effects of $Y_d$ on additional heating. We varied $\mu_{dam}$ from $10^{-4}$ to 0.6. The maximum value of $\mu_{dam}$ is typical for rocky granular media (e.g., Collins et al., 2004).

A thermal softening model (Ohnaka, 1995) was implemented to reproduce the strength behavior as a function of temperature. The yield strength decreases to zero at the melting temperature. The input parameters for the strength model are listed in Supporting Information Table S1.

Gravity was not considered in our calculations because gravitational acceleration is expected to be negligible in the early stages of impact-induced hydrodynamic motion. A uniform temperature of 220 K was assumed for both projectile and target, which corresponds to the typical equilibrium temperature in the main asteroid belt. We only modeled vertical impacts of spherical projectiles onto flat targets, using cylindrical coordinates. We divided a spherical projectile into 50 cells per projectile radius (CPPR). The impact velocity, $v_{imp}$, was varied from 1 to 20 km s$^{-1}$.

Although the projectile radius, $R_p$, was set at 25 km, we can convert the results to any impactor size, because all hydrodynamic equations can be rewritten in dimensionless form in the absence of gravity and for scale-independent strengths (e.g., Johnson and Melosh, 2013). The target was defined as a cylinder with a radius of 24 $R_p$. We followed the simulations until a time $t = 19t_s$, where $t$ and $t_s$ are the time after initial contact between projectile and target, and the characteristic time for projectile penetration ($t_s = 2R_p/v_{imp}$), respectively. The end time corresponds to the initial phase of the excavation stage, when most of the projectile's kinetic energy has already been transferred to the target (e.g., O'Keefe and Ahrens, 1982).

Lagrangian tracer particles were inserted into each computational cell to track the thermal history of each particle. We stored the temporal variations in the spatial position, pressure, and absolute entropy of each tracer particle. Entropy is a better indicator of the energy partitioning into internal degrees of freedom than temperature (e.g., Melosh, 2007). The iSALE model set-up is summarized in Supporting Information Table S2. For comparison, we also conducted simulations without the strength model; i.e., purely hydrodynamic simulations. The results were analyzed using a post-analysis script, which was essentially the same as that employed in previous studies (Nagaki et al., 2016; Kurosawa et al., 2018).

Figure 1 shows the Hugoniot curve for dunite employed here in the entropy–pressure plane. The entropies for incipient ($S_{im}$) and complete melting ($S_{cm}$), as well as that for the reset of the $^{40}Ar$–$^{36}Ar$ age ($S_{Ar}$), are shown as vertical dotted lines. Since the temperatures for such events at a pressure of $10^5$ Pa have already been reported in the literature, we can determine the entropies using the ANEOS. We adopted $S_{Ar}$ = 1.99 kJ K$^{-1}$ kg$^{-1}$, which corresponds to a temperature of 1000 K at $10^5$ Pa, as required for rapid Ar loss (Marchi et al., 2013). We also adopted $S_{im}$ = 2.35 kJ K$^{-1}$ kg$^{-1}$ and $S_{cm}$ = 3.31 kJ K$^{-1}$ kg$^{-1}$, which correspond to the solidus of typical chondritic materials (1373 K at $10^5$ Pa; e.g., Keil et al., 1997) and the liquidus for forsterite, which is the dominant mineral in dunite (2173 K at $10^5$ Pa; e.g., Ahrens and O'Keefe, 1972), respectively. The estimated velocity thresholds for Ar loss and incipient and complete melting are respectively 6.1 km s$^{-1}$ (91 GPa), 6.7 km s$^{-1}$ (105 GPa), and 9.0 km s$^{-1}$ (168 GPa) based on the entropy matching method for head-on collisions between two dunite bodies.

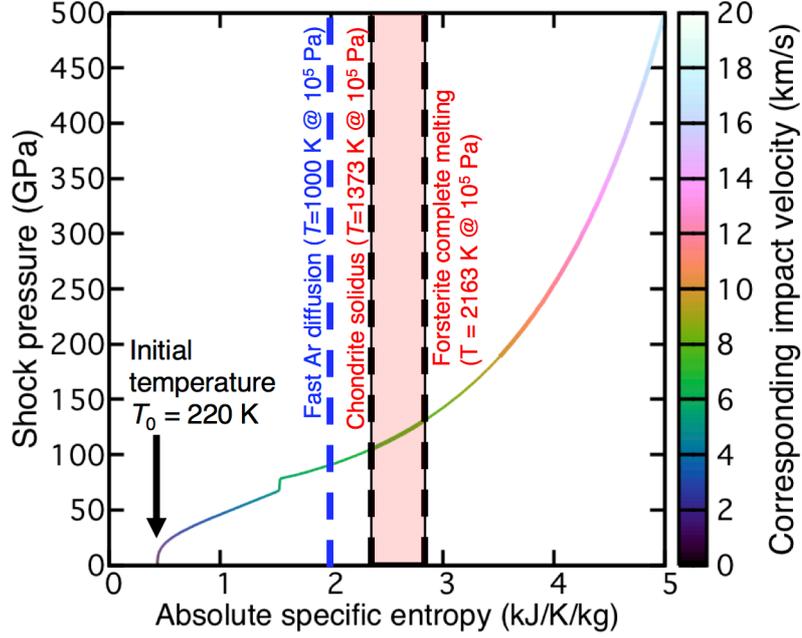

**Figure 1.** Hugoniot curve for dunite used in the computation in the entropy–pressure (*S–P*) plane. This curve was obtained using the ANEOS package. The colors reflect the corresponding impact velocities, calculated as the two-fold particle velocity behind the shock wave. The entropies required for rapid Ar diffusion, incipient melting, and complete melting are shown as the three vertical dashed lines.

We investigated the mass experiencing Ar loss, $M_{Ar}$, as well as the melting mass, $M_{melt}$. Note that we only analyzed $M_{Ar}$ and $M_{melt}$ for the target material. $M_{Ar}$ was calculated as the sum of the tracer mass $m$ for $S > S_{Ar}$; $M_{melt}$ was calculated using the lever rule, which depends on $S$ of each tracer particle, as follows:

$$M_{melt} = M_{melt1} + M_{melt2} + M_{melt3}, \quad (3)$$

where

$$M_{melt1} = \Sigma \frac{S - S_{im}}{S_{cm} - S_{im}} m \quad (S_{im} < S < S_{cm}), \quad (4)$$

$$M_{melt2} = \Sigma m \quad (S_{cm} < S < S_{iv}), \text{ and} \quad (5)$$

$$M_{melt3} = \Sigma \frac{S_{cv} - S}{S_{cv} - S_{iv}} m \quad (S_{iv} < S < S_{cv}), \quad (6)$$

where $S_{iv} = 3.66$ kJ K$^{-1}$ kg$^{-1}$ and $S_{cv} = 7.41$ kJ K$^{-1}$ kg$^{-1}$ are the entropies for the incipient and complete vaporization of forsterite, respectively (Ahrens and O'Keefe, 1972).

Technically speaking, the entropy raised by shock heating tends to slightly decrease during pressure release in grid-based hydrocodes, possibly because of numerical diffusion. Thus, we used the maximum entropy stored in each tracer to calculate $M_{Ar}$ and $M_{melt}$ in both cases, with and without material strength, to extract the degree of entropy enhancement during pressure release, depending on $\mu_{dam}$.

## 3. Results

Figure 2a and 2b shows snapshots from iSALE simulations for vertical impacts at 6 km s$^{-1}$ and $t = 3\ t_s$ without strength and for $\mu_{dam} = 0.6$, respectively. All tracers are colored, depending on their temporal entropy. The trajectories of five selected tracers initially located at $r = 0.76\ R_p$ (where $r$ is the horizontal distance from the impact point) are also shown for reference. Although the shapes of the transient cavities are similar, the temporal entropies for $\mu_{dam} = 0.6$ are much higher than those in the hydrodynamic case, thus showing that the additional heating reported by Quintana et al. (2015) is reproduced by our numerical model.

Figure 2c and 2d shows temporal variations in entropy in the entropy–pressure ($S$–$P$) plane for the iSALE runs shown in Figs 2a and 2b, respectively. Only the $S$–$P$ histories of the tracers initially located at $r = 0.76\ R_p$ are plotted. The data points are colored depending on the tracers' peak pressure. The $S$–$P$ tracks of the five selected tracers in Fig. 2a and 2b are also shown (black symbols). We found that the entropy gradually increases during pressure release in the case of $\mu_{dam} = 0.6$ as shown in Fig. 2d. Most of the entropy increase occurs until the pressure decreases to 3 GPa. Some of the tracers eventually exceed $S_{Ar}$ and $S_{im}$, even though the impact velocity at 6 km s$^{-1}$ is lower than the velocity thresholds obtained from the entropy matching method. In contrast, Fig. 2c shows the entropy is mostly preserved during pressure release when we did not implement the strength model. Note that the peak pressures of the tracers in Fig. 2d deviate from the Hugoniot curve, because here we plot the 'mean pressure,' not the longitudinal stress of elastoplastic media, for which the Hugoniot relations hold strictly. The mean pressure is always less than the longitudinal stress by a factor that depends on the Poisson ratio. In contrast, Fig. 2c shows that the tracers mostly follow the Hugoniot curve immediately after the shock wave's arrival in the hydrodynamic case. These results indicate that iSALE reproduces the hydrodynamic and elastoplastic behaviors accurately. Figure 3a and 3b shows $M_{Ar}$ and $M_{melt}$ as functions of internal friction and impact velocity, respectively. We found that $M_{Ar}$ and $M_{melt}$ increase significantly beyond $\mu_{dam} = 10^{-2}$ at relatively low $v_{imp}$. On the other hand, if $\mu_{dam} < 10^{-2}$, regardless of $\mu_{dam}$, $M_{Ar}$ and $M_{melt}$ are nearly constant and similar to the equivalent masses in the purely hydrodynamic case. This means that any additional heating caused by $Y_i$ and $Y_{coh}$ in Eqs. (1) and (2) is not significant. The rates of increase in $M_{Ar}$ and $M_{melt}$ seems to decrease for $\mu_{dam} > 0.1$, suggesting that the effect of friction on the additional heating becomes less significant. Plastic deformation versus the limiting strength plays a major role on the additional heating for $\mu_{dam} > 0.1$ (See also Supporting information S1). The $\mu_{dam}$ dependence on $M_{Ar}$ and $M_{melt}$ also tends to reduce as $v_{imp}$ increases, as thermal softening occurs at high $v_{imp}$. This finding is also consistent with the results of Quintana et al. (2015). Although we only modeled vertical impacts, we can approximately estimate the masses for oblique impacts because the degree of shock heating correlates well with the normal component of the impact velocity (Pierazzo and Melosh, 2000). The converted impact velocities, at a 45° angle measured from the tangent plane, are indicated in parentheses next to the color bar in the figure.

Next, we discuss the threshold pertaining to the impact velocities for Ar loss and incipient melting, $v_{Ar}$ and $v_{melt}$, respectively. If we define the velocity thresholds when $M_{Ar}$ and $M_{melt}$ exceed 1 wt% of the projectile mass, the resulting thresholds in the hydrodynamic case are 6 and 8 km s$^{-1}$, respectively. These values are close to the prediction of the entropy-matching method mentioned in Section 2. In the case for $\mu_{dam} = 0.6$, $v_{Ar}$ and $v_{melt}$ largely decrease to 1.5 and 4 km s$^{-1}$, respectively.

Here as a conservative estimate, we also estimated the velocities when $M_{Ar}$ and $M_{melt}$ exceed 10 wt% of the projectile mass as the velocity thresholds in this study. This is because local energy concentrations owing to a few processes, such as jetting (e.g., Kieffer, 1977; Sugita and Schultz, 1999; Johnson et al., 2015; Kurosawa et al. 2015) and shear banding (Kondo and Ahrens, 1983), are expected to become more important than the bulk shock heating considered here when we deal with a small fraction of the heated mass. In addition, numerical simulations often cause overshooting of the temperature over a range of 3–10 cells near the contact boundary between a moving projectile and a target. The mass under such artificial overheated conditions would reach 1−10 wt% of the projectile. When we adopt 10 wt% of the projectile mass for the incipient melting condition, the estimated velocity thresholds $v_{Ar}$ and $v_{melt}$ are 8 and 10 km s$^{-1}$, respectively, for the purely hydrodynamic model. For $\mu_{dam} = 0.6$, $v_{Ar}$ and $v_{melt}$ show significant decrease to 2 and 6 km s$^{-1}$, respectively.

We also investigated the effects of varying the spatial resolution on the heated mass (Supporting Information S3). We confirmed that the results converge to nearly the same value in the CPPR range from 25 to 200.

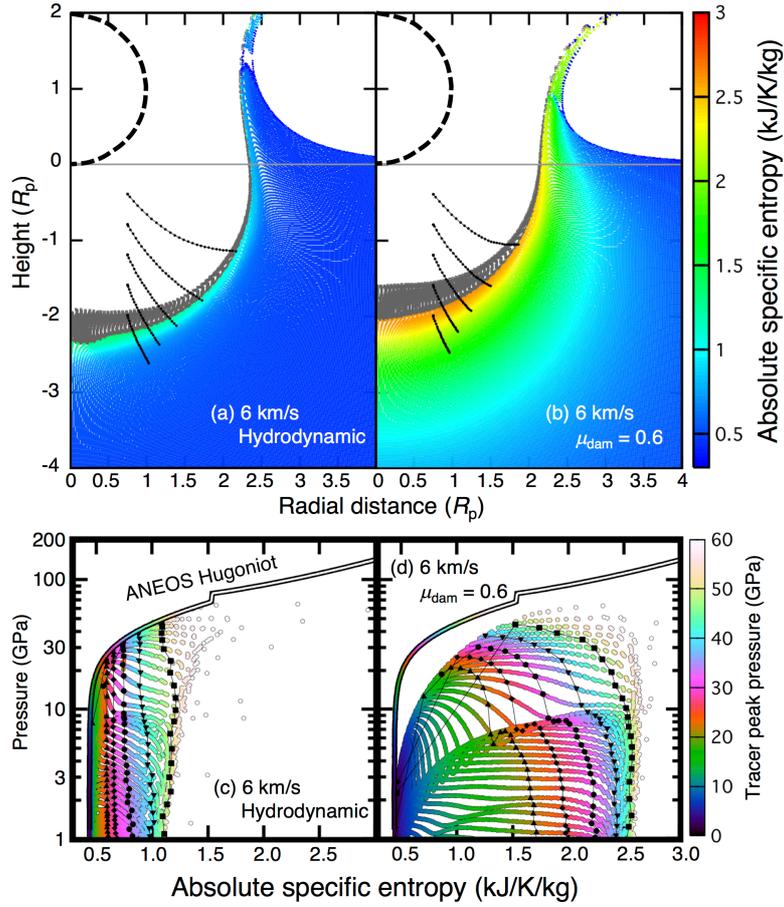

**Figure 2.** Snapshots of the iSALE simulations for a vertical impact at 6 km s$^{-1}$ and $t = 3$ $t_s$ (upper panels) and thermodynamic tracks of selected tracer particles in the $S$–$P$ plane (lower panels) from the initial contact to $t = 19$ $t_s$. The case without strength is shown in (a) and (c), and the case with strength ($\mu_{dam} = 0.6$) is shown in (b) and (d). The tracer trajectories of the five selected tracers are also shown as black filled circles. In lower panels, thermodynamic tracks for tracer particles initially located at a horizontal distance of 0.76 $R_p$ are only plotted.

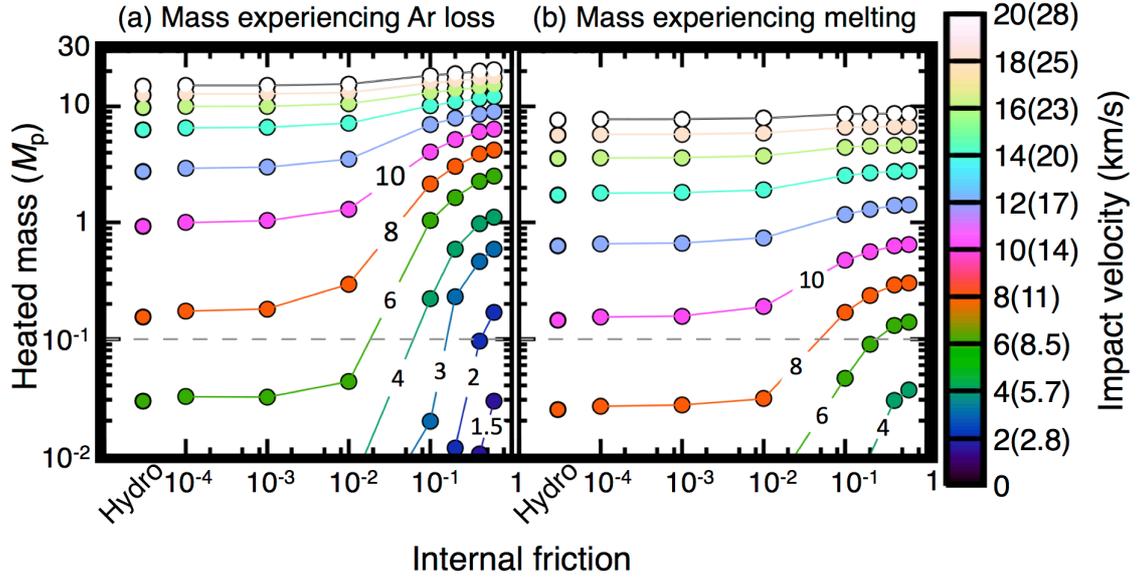

**Figure 3.** Mass experiencing Ar loss $M_{Ar}$ (a) and experiencing melting $M_{melt}$ (b) as a function of internal friction and impact velocity. The results for the hydrodynamic case are also shown (left in each panel) for comparison. The data points are colored according to their impact velocity. The numbers on the lines indicate the impact velocities. Values in parentheses indicate the impact velocities for oblique impacts at 45° from the tangent plane.

## 4. Discussion

Here, we discuss the reasons why the additional heating is obvious when $\mu_{dam} > 10^{-2}$. The energy required to move materials supported by strength, herein referred to as the 'specific strength energy,' $e_{strength}$, is approximately expressed as

$$e_{strength} = \varepsilon Y_d / \rho \ [\text{J kg}^{-1}], \quad (7)$$

where $\rho \sim 3000$ kg m$^{-3}$ and $\varepsilon$ are the density and volumetric strain, respectively. This energy corresponds to the energy converted from kinetic to internal energy in damaged rocks under a given pressure $P$. Strictly speaking, $e_{strength}$ should be an integral of the stress times the strain over time. Here, we assumed that strain is a constant and is equal to unity for an order-of-magnitude estimate. The expected rise in temperature, $\Delta T$, is approximately expressed as

$$\Delta T = e_{\text{strength}}/C_p \; [\text{K}], \tag{8}$$

where $C_p \sim 1000$ J K$^{-1}$ kg$^{-1}$ is the isobaric specific heat. Figure 4 shows $\Delta T$ as a function of both $\mu_{\text{dam}}$ and $P$. We used Eq. (S2) in Supporting Information S1 here. If we focus on pressures $P$ around 10 GPa (the typical peak pressure for an impact at several km s$^{-1}$ is ~100 GPa), where a large entropy increase is observed in the simulation (see Fig. 2d), $\Delta T$ exceeds 100 K for $\mu_{\text{dam}} > 10^{-2}$. For typical rocks ($\mu_{\text{dam}} = 0.6$), $\Delta T \sim 1000$ K. The change in entropy, $\Delta S$, is approximately expressed as

$$\Delta S = C_p \ln(1 + \Delta T/T_{\text{before}}) \; [\text{J K}^{-1} \text{kg}^{-1}], \tag{9}$$

where $T_{\text{before}}$ is the temporal temperature before heating occurs. If we consider the situation for $\Delta T = 1000$ K and $T_{\text{before}} = 1200$ K, $\Delta S$ is approximately 1 kJ K$^{-1}$ kg$^{-1}$. This order-of-magnitude estimate is fully consistent with the calculation results (Fig. 2).

It is necessary to assess the consistency in energy partitioning. The kinetic energy in the damaged materials should be converted into internal energy to explain the additional heating. We confirmed that the kinetic energy lost owing to the material strength accounts for the additional heating during pressure release by examining the kinetic energy carried away by the ejecta (Supporting information S4).

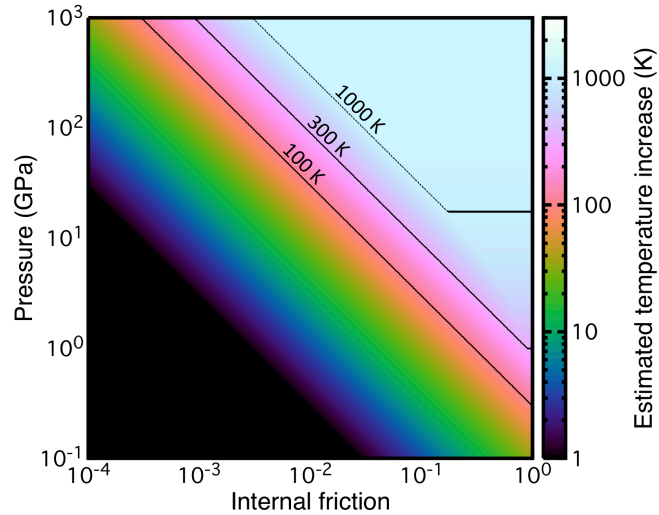

**Figure 4.** Expected temperature rise, $\Delta T$, as function of internal friction in damaged rocks, $\mu_{\text{dam}}$, and temporal pressure, $P$. Three isolines for 100, 300, and 1000 K are shown.

We have demonstrated that the impact-induced reset of the $^{40}$Ar–$^{36}$Ar age and melting are much easier to achieve than previously thought, owing to energy conversion from kinetic to internal energy in impact-comminuted damaged rocks. We used a different numerical code to that employed in previous studies, and the results strongly indicate that the additional heating reported by Quintana et al. (2015) is physically real. The velocity thresholds for various geological materials, as obtained from application of the entropy matching method in previous studies, would need to be significantly revised compared with those based on using shock-physics codes combined with constitutive models. In addition, our results imply that the choice and accuracy of constitutive models included in shock-physics codes have a marked effect on the degree of shock heating in numerical models.

## 5. Conclusions

We investigated the effects of material strength on the degree of impact heating using analytical and numerical approaches. We numerically investigated the influence of internal friction in damaged rocks, $\mu_{dam}$, on the degree of impact-induced heating for vertical impacts at 1–20 km s$^{-1}$. The additional heating during pressure release in targets with strength becomes obvious for $\mu_{dam} > 10^{-2}$, which agrees well with simple analytical considerations. Our numerical results demonstrate that the impact velocity required for the reset of $^{40}$Ar–$^{36}$Ar ages and incipient melting of the target with the mass of 10 wt% of the projectile mass is reduced from 8 to 2 km s$^{-1}$ and 10 to 6 km s$^{-1}$, respectively, when material strength is considered.

**Acknowledgements**

We thank the developers of iSALE, including G. Collins, K. Wünnemann, B. Ivanov, J. Melosh, and D. Elbeshausen. The 'quick look' of the iSALE results using the pySALEPlot tool written by Tom Davison greatly helped us to conduct the series of numerical simulations. We acknowledge useful discussions at a workshop on planetary impacts held at Kobe University, Japan, in 2016 and at the JpGU–AGU Joint meeting in 2017. We also thank Boris A. Ivanov and H. Jay Melosh for their constructive reviews that helped greatly improve the manuscript, and Andrew J. Dombard for handling the manuscript as an editor. The authors (KK and HG) are supported by JSPS KAKENHI Grant No. JP17H02990. KK is supported by JSPS KAKENHI Grant Nos JP17K18812,

JP17H01176, and JP17H01175. HG is also supported by MEXT KAKENHI Grant No. JP17H06457. The iSALE shock physics code is not fully open-source: it is distributed on a case-by-case basis to academic users in the impact community, for noncommercial use only. A description of the application requirements can be found at the iSALE website (http://www.isale-code.de/redmine/projects/isale/wiki/Terms_of_use). Calculated results are available by contacting the corresponding author (kosuke.kurosawa@perc.it-chiba.ac.jp). Supporting information can be found in the online version of the manuscript.

hydrodynamic code III: Revised analytical equation of state, pp. *SC-RR-71 0714* **119** pp., Sandia Laboratories, Albuquerque, NM.

Wünnemann, K., G. S. Collins, and H. J. Melosh (2006), A strain-based porosity model for use in hydrocode simulations of impacts and implications for transient crater growth in porous targets, *Icarus*, **180**, 514–527.

Wünnemann, K., G. S. Collins, and G. R. Osinski (2008), Numerical modelling of impact melt production in porous rocks, *Earth and Planetary Science Letters*, **269**, 530-539.
16

*Geophysical Research Letters*

Supporting Information for

**Effects of friction and plastic deformation in shock-comminuted damaged rocks on impact heating**

Kosuke Kurosawa[1]* and Hidenori Genda[2]

[1]Planetary Exploration Research Center, Chiba Institute of Technology, 2-17-1, Narashino, Tsudanuma, Chiba 275-0016, Japan

[2]Earth–Life Science Institute, Tokyo Institute of Technology, 2-12-1 Ookayama, Meguro-ku, Tokyo 152-8550, Japan

**Contents of this file**

> Text S1 to S4
> Figure S1 to S3
> Tables S1 and S2

**Introduction**

This document includes the input parameters pertaining to the constitutive model employed in this study (Text S1; Figure S1; Table S1), the iSALE model setup (Text S2; Table S2), the effect of spatial resolution on the results (Text S3, Figure S2), and the effect of strength on the kinetic energy carried away by the ejecta (Text S4, Figure S3).

**Text S1.**

We employed an experiment-based strength model to treat elastoplastic behavior of rocky materials (Eq. 1 in the Section 2 of the Main text). The pressure-dependent yield strength for intact rock $Y_i$ is approximate by (Lundborg, 1968)

$$Y_i = Y_{coh,i} + \frac{\mu_{int} P}{1 + \frac{\mu_{int} P}{Y_{limit} - Y_{coh,i}}}, \quad (S1)$$

where $Y_{coh,i}$, $\mu_{int}$, $P$, and $Y_{limit}$ are the cohesion for intact rock at zero pressure, the coefficient of internal friction for intact rock, temporal mean pressure, and the limiting strength at high pressure, respectively. The limiting strength $Y_{limit}$ is known as the von Mises Plastic limit. Since the yield strength for comminuted rocks $Y_d$ is also limited by the plastic limit, the expression of $Y_d$ actually used in our model is

$$Y_d = \min(Y_{coh} + \mu_{dam} P, Y_i), \quad (S2)$$

where $Y_{coh}$ and $\mu_{dam}$ are the cohesion in damaged rocks and the coefficient of internal friction for damaged rocks, respectively. We list the input parameters for the constitutive model in Table S1. Figure S1 shows the yield strength $Y_d$ as a function of mean pressure with different values of $\mu_{dam}$. The slopes of the straight lines correspond to $\mu_{dam}$. At lower pressures than the intersects between the straight lines for damaged rock and the curve for intact rock, the internal friction plays main role to produce the additional heat. Above the pressures, the plastic deformation against the limiting strength becomes more important.

**Table S1.** Input parameters for the strength model.

| | |
|---|---|
| EOS type | ANEOS[a] |
| Material | Dunite |
| Strength model | Rock[b] |
| Poisson ratio | 0.25[c] |
| Melting temperature (K) | 1373[c] |
| Thermal softening coefficient | 1.1[c] |
| Simon parameter A (GPa) | 1.52[c] |
| Simon parameter C | 4.05[c] |
| Cohesion (undamaged) (MPa), $Y_{coh,i}$ | 10[d] |
| Cohesion (damaged) (kPa), $Y_{coh}$ | 10[d] |
| Internal friction (undamaged), $\mu_{int}$ | 1.2[d] |
| Internal friction (damaged), $\mu_{dam}$ | $10^{-4}$ to 0.6[e] |
| Limiting strength (GPa), $Y_{limit}$ | 3.5[d] |
| Minimum failure strain | $10^{-4}$ [d] |
| Constant for the damage model | $10^{-11}$ [d] |
| Threshold pressure for the damage model (MPa) | 300[d] |

a. Benz et al. (1989)

b. A detailed description of the strength model for rocks can be found in Collins et al. (2004).

c. Johnson et al. (2015)

d. Typical values for minerals are employed.

e. The maximum value is taken from Johnson et al. (2015).

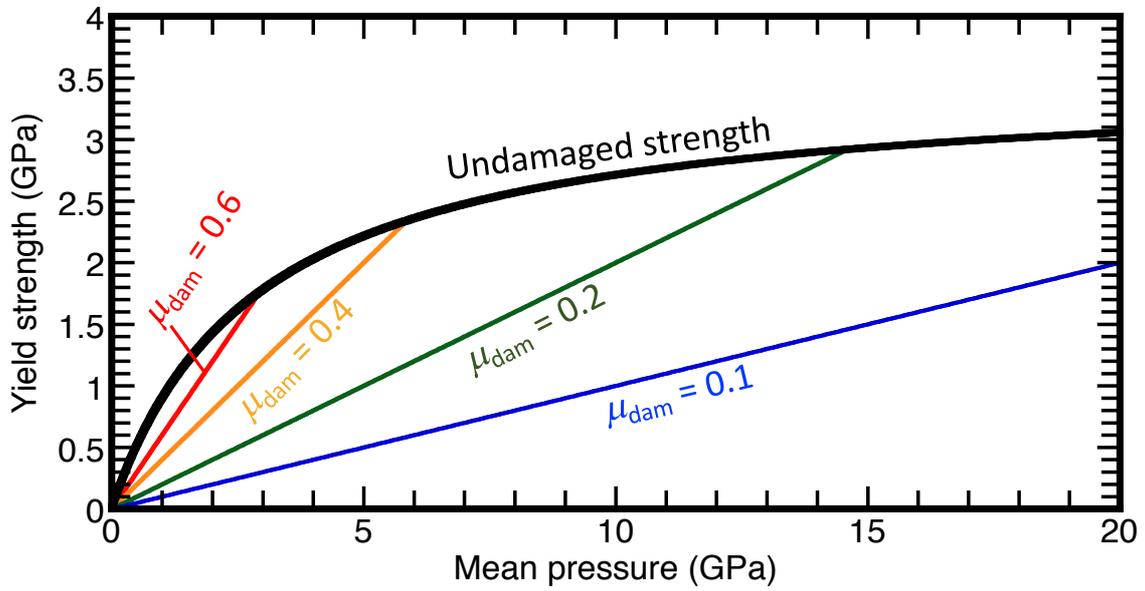

**Figure S1.** Yield strength for damaged rock as a function of mean pressure with different $\mu_{dam}$ = 0.1(blue), 0.2(green), 0.4(orange), and 0.6(red). The yield strength for intact rock given by Eq. (S1) is plotted as thick black curve.

**Text S2.**

We describe the iSALE model setup in Table S2. The full description for each value can be found in the iSALE manual (Collins et al., 2016).

**Table S2.** Input parameters for the 2D iSALE calculations.

| | |
|---|---|
| Computational geometry | Cylindrical coordinates |
| Number of computational cells in the $R$ direction | 1200 |
| Number of computational cells in the $Z$ direction | 1500 |
| Number of cells for extension in the $R$ direction | 200 |
| Number of cells for extension in the $Z$ direction (bottom) | 200 |
| Number of cells for extension in the $Z$ direction (top) | 100 |
| Extension factor | 1.02 |
| Cells per projectile radius (CPPR)[b] | 50 |
| Grid spacing (m/grid) | 500 |
| Artificial viscosity $a_1$ | 0.24 |
| Artificial viscosity $a_2$ | 1.2 |
| Impact velocity (km s$^{-1}$) | 1, 1.5, 2, 3, 4, 6, 8, 10, 12, 14, 16, 18, 20 |
| High-speed cutoff | two-fold impact velocity |
| Low-density cutoff (kg m$^{-3}$) | 1 |

**Text S3.**

We conducted another series of iSALE runs to address the effects of differences in spatial resolution. We varied the number of cells per projectile radius from 25 to 200 in this series. The impact velocity was fixed at 6 km s$^{-1}$. The other input values in the iSALE model were the same as for the regular runs. Figure S2 shows $M_{Ar}$ and $M_{melt}$ as a function of the scaled time $t/t_s$, where $t_s = 2R_p/v_{imp}$, $R_p$, and $v_{imp}$ are the characteristic time for projectile penetration, the projectile radius, and the impact velocity, respectively. Although $M_{Ar}$ and $M_{melt}$ tend to show larger values for lower CPPR values during the early stages of impact events, $t/t_s < 1$, they converge to similar values by $t/t_s = 10$. Thus, the spatial resolution employed in this study is sufficiently high to investigate $M_{Ar}$ and $M_{melt}$.

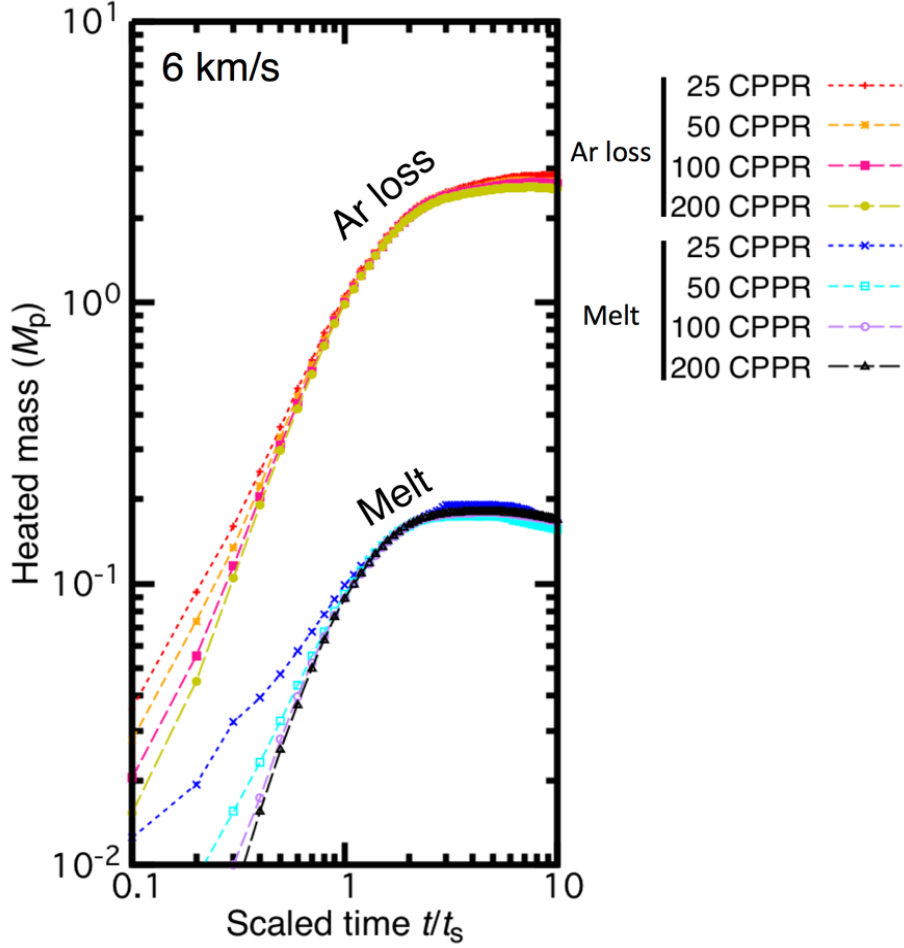

**Figure S2.** Heated masses $M_{Ar}$ and $M_{melt}$ as a function of scaled time. The masses $M_{Ar}$ and $M_{melt}$ for different values are shown using different lines and colors.

**Text S4.**

Figure S3 shows the cumulative ejected mass, $M_{ej}$, for velocities higher than a given velocity prior to $t = 19t_s$ as a function of ejection velocity, $v_{ej}$. We plotted the results for 6 km s$^{-1}$ including material strength for $\mu_{dam} = 0.6$ (red) and without strength (blue). The ejected mass, $M_{ej}$, for $\mu_{dam} = 0.6$ is systematically smaller than for the hydrodynamic case. The difference in $M_{ej}$ between the cases is a factor of 2–4. The difference in the kinetic energy carried away by the ejecta is also shown, along the right-hand y axis (black), showing that it reaches ~2 MJ kg$^{-1}$. In this calculation, the cumulative kinetic energy carried away by the ejecta is divided by the projectile mass. Since $C_p$ ~1000 J K$^{-1}$ kg$^{-1}$, this energy difference corresponds to a temperature increase of ~2000 K in rocky materials with a mass of 1 $M_p$. Consequently, the kinetic energy lost owing to the material strength accounts for the additional heating during pressure release.

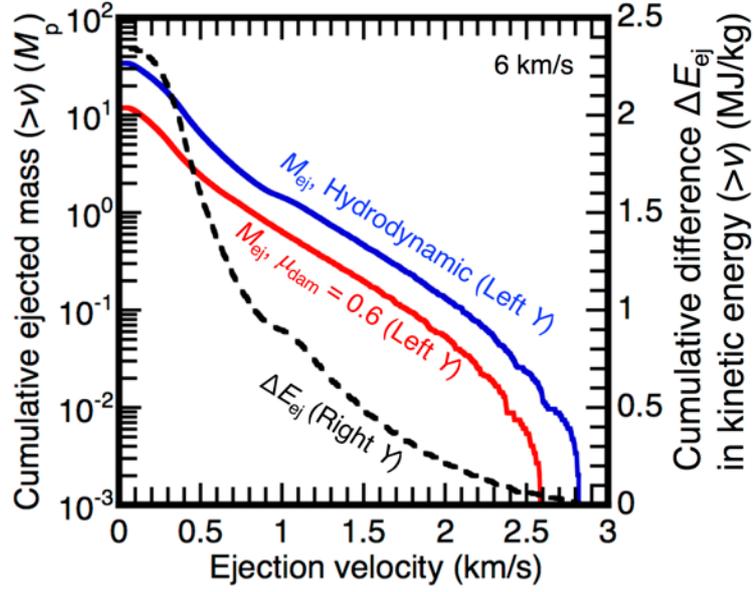

**Figure S3.** Cumulative mass ejected for velocities in excess of a given ejection velocity prior to $t = 19\ t_s$ and as a function of ejection velocity. Results including material strength for $\mu_{dam} = 0.6$ and without strength are shown as red and blue lines (left-hand $y$ axis), respectively. The difference in the kinetic energy carried away by the ejecta between the cases is shown as the dashed black line (right-hand $y$ axis).